\documentclass[12pt]{article}

\usepackage[]{graphicx,bm,color,subfigure,amsmath}

\usepackage{scicite}

\usepackage{times}
\usepackage{textcomp}
\usepackage{gensymb}

\newenvironment{sciabstract}{%
\begin{quote} \bf}
{\end{quote}}

\newcounter{lastnote}

\title{An Intrinsic Spin Orbit Torque Nano-Oscillator}

\author
{M. Haidar,$^1$ A. A. Awad,$^1$ M. Dvornik,$^1$ R. Khymyn,$^1$ A. Houshang,$^1$ \\ J. \AA kerman$^{1,2\ast}$\\ 
\\
\normalsize{$^{1}$Physics Department, University of Gothenburg, 412 96 Gothenburg, Sweden}\\
\normalsize{$^{2}$Material Physics, School of Engineering Sciences,}\\ 
\normalsize{KTH Royal Institute of Technology, Electrum 229, 164 40 Kista,Sweden}\\
\\
\normalsize{$^\ast$To whom correspondence should be addressed; E-mail:  johan.akerman@physics.gu.se.}
}
\date{}

\begin{document} 


\baselineskip24pt


\maketitle


\begin{sciabstract}
 
Spin torque and spin Hall effect nano-oscillators 
generate high intensity spin wave auto-oscillations on the nanoscale enabling novel microwave applications in spintronics, magnonics, and neuromorphic computing. 
For their operation, these devices require externally generated spin currents 
either from an additional ferromagnetic layer or a material with a high spin Hall angle. Here we demonstrate highly coherent field and current tunable microwave signals 
from nano-constrictions in single 15--20 nm thick permalloy layers. Using a combination of spin torque ferromagnetic resonance measurements, scanning micro-Brillouin light scattering microscopy, and micromagnetic simulations, we identify the auto-oscillations as emanating from a localized edge mode of the nano-constriction driven by spin-orbit torques. Our results pave the way for greatly simplified designs of auto-oscillating nano-magnetic systems only requiring a single ferromagnetic layer.


\end{sciabstract}


\section*{Introduction}

Spin transfer torque\cite{berger1996prb,slonczewski} (STT)---the transfer of angular momentum from a spin current to the local spins---can act as negative spin wave damping and drive the magnetization of a ferromagnet (FM) into a state of sustained large-angle precession
For STT to occur, one first needs to generate a spin current and then inject this spin current into a ferromagnetic layer with a magnetization direction that is non-collinear with the polarization axis of the spin current. In so-called spin torque nano-oscillators (STNOs) these conditions can be fulfilled 
either in giant magnetoresistance trilayers, or in magnetic tunnel junctions, 
if the two ferromagnetic layers are in a non-collinear state \cite{Chen2016}. More recently, the spin Hall effect \cite{Hirsh1999,D'yakonov1971, Sinova2015} was instead used to create pure spin currents injected into an adjacent ferromagnetic layer, in devices known as spin Hall nano-oscillators (SHNOs) \cite{Chen2016}. These are both much easier to fabricate \cite{Demidov2014apl}, and exhibit superior synchronization properties and therefore much higher signal coherence \cite{Awad2017}. Even so, SHNOs still suffer from a number of drawbacks and conflicting requirements: \emph{i}) the non-magnetic layer generating the spin current dramatically increases the zero-current spin wave damping of the ferromagnetic layer (up to 3x for NiFe/Pt \cite{Nanprb2015}), \emph{ii}) the surface nature of the spin Hall effect requires ultrathin ferromagnetic layers for reasonable threshold currents, which further increases the spin wave damping, and \emph{iii}) since the current is shared between the driving layer and the magnetoresistive ferromagnetic layer, neither driving nor signal generation benefits from the total current, leading to non-optimal threshold currents, low output power, unnecessary dissipation, and heating. It would therefore be highly advantageous if one could do away with the additional metal layers for operation.



It was recently realized that spin-orbit torque can emerge 
from Berry curvature due to a broken inversion symmetry\cite{kurebayashi2014antidamping}, either in the bulk of the magnetic material\cite{fang2011spin,Safranski2017planar} or at its interfaces. In particular, interfaces to insulating oxides\cite{Miron2011,an2016spin,Emori2016prb,demasius2016enhanced,Gao2018prl} can lead to a non-uniform spin distribution over the film thickness \cite {haidar2013prb,Tsukahara2014prb,Azevedo2015prb}. As oxide spin-orbitronics has seen considerable development in recent years\cite{Vaz2018jjap, manchon2018current}, oxide based spin-orbit torque nano-oscillators with a single metallic layer could be envisioned. If so, the current would be confined to the ferromagnetic metal layer, avoiding the detrimental current sharing between adjacent metal layers and leading to an overall lower power consumption.


Here we propose, and experimentally demonstrate, an intrinsic spin orbit torque (SOT) nano-oscillator based on a single permalloy layer with thickness in the 15--20 nm range, grown on an Al$_2$O$_3$ substrate and capped with SiO$_2$. The magnetodynamics of the unpatterned stacks is characterized using ferromagnetic resonance (FMR) spectroscopy and the linear spin wave modes of the final devices are determined using spin-torque FMR. Spin wave auto-oscillations are observed both electrically, via a microwave voltage from the anisotropic magnetoresistance, and optically, using micro-Brillouin Light Scattering ($\mu$-BLS) microscopy. The current-field symmetry of the observed auto-oscillations agree with the expected symmetry for spin orbit torques, 
either intrinsic to the Py layer itself or originating from the Al$_2$O$_3$/Py and Py/SiO$_2$ interfaces. 

Fig.~1A shows a scanning electron microscopy image of a $w$ = 30 nm wide nano-constriction etched out of $t$ = 15 nm thick single permalloy (Ni$_{80}$Fe$_{20}$; Py) layer, magnetron sputtered onto a sapphire substrate. 
On top of the nano-constriction a coplanar waveguide is fabricated for electrical measurements (not shown). During operation, an external magnetic field $(H)$, applied at an out-of-plane angle $\theta$ = $75^{\circ}$ and an in-plane angle $\phi$ = $22^{\circ}$, tilts the Py magnetization out of plane. 

Fig.~1B shows the thickness dependence of the Gilbert damping ($\alpha$) of the unprocessed Py films as well as the resistance (R) and the anisotropic magnetoresistance (AMR) of the final devices (inset). As the thickness of the Py layer increases, we measure a noticeable decrease of $\alpha$ and R, and a corresponding increase of AMR \cite{Ingvarssonprb2002}. Thus, using thicker permalloy films could be advantageous as compared to thinner films where by reducing $\alpha$ one can reduce the threshold current for driving spin wave auto-oscillations and by increasing AMR one can achieve larger output power as the electrical read-out is based on AMR in these devices.

\begin{figure}
\begin{center}
\includegraphics[width=0.8\textwidth]{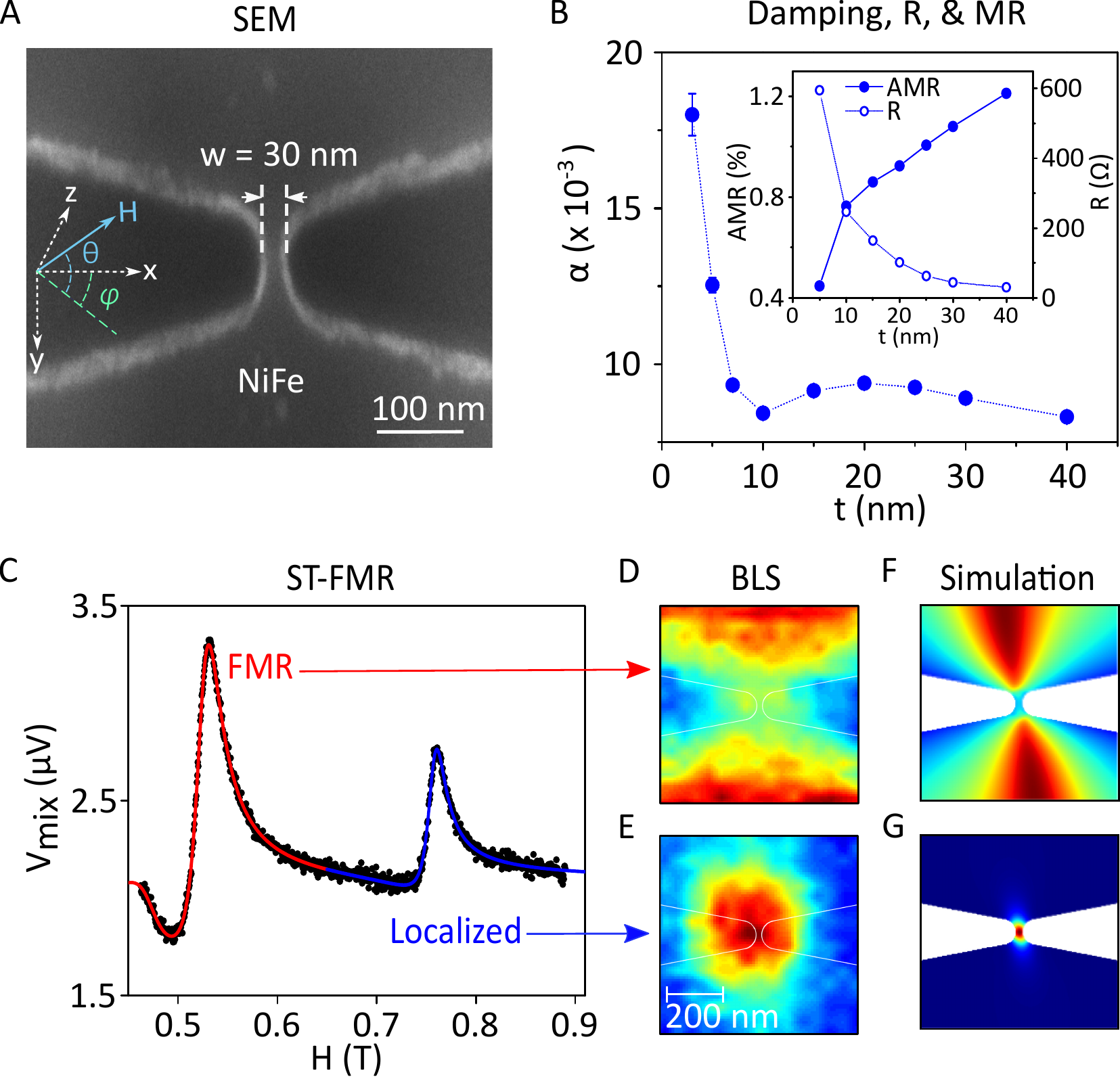}
\caption{\textbf{Nano-constriction fabrication, characterization and simulation.} (\textbf{A}) Scanning electron microscope image of a permalloy nano-constriction 
showing its width ($w$ = 30 nm) and the direction of the applied field. (\textbf{B}) Gilbert damping of blanket permalloy films as a function of film thickness. Inset: AMR (closed symbol) and resistance (open symbol) vs.~thickness of 30 nm wide nano-constrictions. (\textbf{C}) ST-FMR spectrum vs.~field at $\it{f}$ = 10 GHz for a $w$ = 30 nm and $t$ = 15 nm nano-constriction showing its two linear modes; 
red and blue lines are Lorentzian fits to the FMR mode and the localized mode, respectively. The spatial distribution of the FMR and the localized modes, imaged by $\mu$-BLS (\textbf{D,E}), and calculated using micromagnetic simulations (\textbf{F,G}).}
\end{center}
\end{figure}

To determine the magnetodynamic properties of the nano-constrictions, we first characterized them in the linear regime using spin torque ferromagnetic resonance spectroscopy (ST-FMR) \cite{Liu2011} in a field applied at $\theta$ = $75^{\circ}$ and $\phi$ = $22^{\circ}$. 
A typical ST-FMR spectrum 
as a function of field magnitude, and at constant frequency \textit{f} = 10 GHz, reveals 
two well-separated resonances that can both be accurately fit with Lorentzians (Fig. 1C). 
The frequency vs.~field dependence of the lower field peak agrees perfectly with the FMR dispersion at the specific angles used, from which we can extract an effective magnetization $\mu_{0}M_{eff}$ = 0.72 T. 
We can directly image the spatial distribution of the two linear modes using micro-Brillouin light scattering ($\mu$-BLS) microscopy of the thermally excited SWs
: the amplitude of the FMR mode is the largest well outside the nano-constriction region (Fig. 1D) whereas the amplitude of the higher-field (lower energy) mode is localized to the center of the nano-constriction (Fig. 1E). The localized mode is confined to the nano-constriction due to a local reduction of the internal magnetic field in this region. The nature of the observed modes is further corroborated by carrying out micromagnetic simulations where we calculate the spatial profile of the magnetization within the nano-constriction. The spatial maps for the simulated FMR mode and the localized mode (Fig. 1F, G) closely agree  with the $\mu$-BLS maps.

We then investigated the impact of a direct current on the linewidths ($\Delta H$) of both the FMR and the localized modes (Fig. 2A). While the linewidths of both modes decreases with current magnitude for an even current-field symmetry 
(the orientation of the field and current satisfy $\mathbf{H} \cdot \mathbf{I} >0$) and increases for an odd symmetry ($\mathbf{H} \cdot \mathbf{I} <0$), the localized mode (blue circles) is three times more strongly affected than the FMR mode. 
This is again consistent with the localized mode residing in the nano-constriction region where the current density is the largest. Naively extrapolating the $\Delta H$ vs.~current data to zero linewidth then predicts that auto-oscillations on this mode should be possible at a current magnitude of 2 mA.

That auto-oscillations are indeed possible to achieve at such currents is experimentally confirmed by sweeping the field magnitude from in Fig.~2B\&C. 
The electrical microwave power spectral density (PSD) is measured at constant currents of $I$ = 2 mA (Fig.~2B) and $I$ = --2 mA (Fig.~2C)  as the field is swept from 0.8 T to -0.8 T. 
Auto-oscillations are detected under positive applied fields between +1 kOe and +8 kOe with a frequency tunable from 3.5 to 12 GHz. Somewhat unexpectedly, a measurable PSD is also observed under \emph{negative} fields between -1 kOe and -3 kOe where the frequency increases from 3.5 to 5.5 GHz. Then, by switching the direction of the current, (Fig. 2C), we measure a corresponding continuous branch of auto-oscillations under negative fields between -1 kOe and -8 kOe and, again, a noticeable PSD appears under a \emph{positive} field between +1 kOe and +3 kOe, this time at an essentially constant frequency of about 4 GHz.

\begin{figure}
\begin{center}
\includegraphics[width=\textwidth]{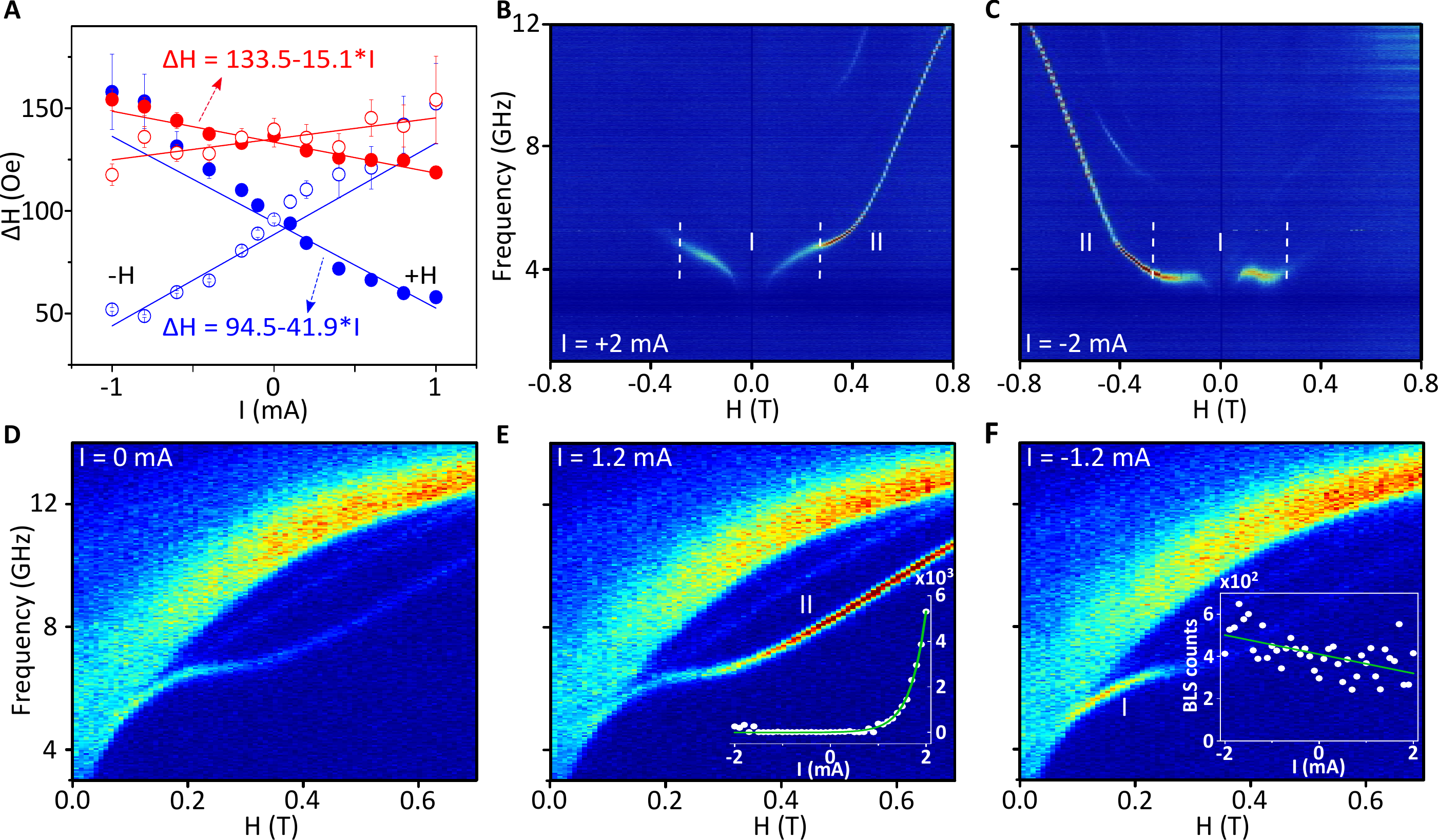}
\caption{\textbf{Current induced spin wave auto-oscillations.} (\textbf{A}) Current dependent ST-FMR linewidth of the FMR (red circles) and localized (blue circles) modes for positive field (filled symbols) and negative field (open symbols). 
(\textbf{B, C}) Electrical measurements of the power spectral densities of spin wave auto-oscillations vs.~applied field measured under a direct current of (\textbf{B}) +2 mA and (\textbf{C}) -2 mA. Label I and II indicate low-field and high-field auto-oscillations, respectively. (\textbf{D, E, F}) $\mu$-BLS measurements of the frequency vs.~applied field showing the intensity of the FMR and the localized mode measured at 0 mA (\textbf{D}), +1.2 mA (\textbf{E}), and -1.2 mA (\textbf{F}). The insets in (\textbf{E, F}) show the change of the BLS counts as a function of current at $\mu_0 H$ = 0.7 T and $\mu_0 H$ = 0.2 T, respectively.}
\end{center}
\end{figure}

We categorize the measured PSD into two regions depending on the strength of the magnetic field: region (I) for low fields $|H| \leq 3$ kOe and region (II) for high fields $|H| > 3$ kOe.
In region (I) the PSD 
is independent of the field-current symmetry, \emph{i.e.}~a measureable PSD is always present regardless of the field and current polarities. The microwave signal in this region is rather weak with a large linewidth. 
In region (II) the auto-oscillations are both sharper and more powerful and are detected only when the orientation of the field and current satisfy $\mathbf{H} \cdot \mathbf{I} >0$. In this region, the spin current creates a spin-torque that is aligned either anti-parallel ($\mathbf{H} \cdot \mathbf{I} >0$) or parallel ($\mathbf{H} \cdot \mathbf{I}<0$) to the damping and hence it either suppresses or enhances the damping depending on symmetry.

To better understand the origin of the two different regions and their distinctly different symmetries we performed $\mu$-BLS measurements vs.~field and current. Fig.~2D shows the frequency vs.~field of the thermally populated linear modes at $I$ = 0 mA, where we observe the localized mode and the continuous SW band above the FMR frequency. When we increase the current to 1.2 mA (Fig.~2E) a noticeable increase in the intensity of the localized mode can be observed for high positive fields, \emph{i.e.} in region (II). It is noteworthy that the SW amplitude in this region depends exponentially on current magnitude, as shown in the inset of (Fig.~2E), consistent with spin-torque driven auto-oscillations. In contrast, by switching the current to --1.2 mA the intensity of the localized mode in region (II) becomes weaker than at zero current, consistent with additional damping from spin-torque, while it increases slightly a low fields (region (I)). The current dependence in region (I) is however much weaker and essentially linear, suggesting thermal activation, 
but not auto-oscillations. We can hence conclude from the $\mu$-BLS characterization that the observed auto-oscillations originate in a continuous manner from the localized mode inside the nano-constriction and that the auto-oscillations require a certain magnetic state reached only for a field above 0.3 T. These results were reproducible from device to device and auto-oscillations were observed for both $t$ = 15 and 20 nm. 

Next, we discuss the possible mechanism that could be driving the auto-oscillations, \emph{i.e.}~for which, the spin current should exert an opposite torque to the Gilbert damping with the observed current-field symmetry. 
Since our devices have a strongly non-uniform magnetization distribution in the vicinity of the constriction and should have a relatively strong temperature gradient during operation, we might have anti-damping originating from \emph{(i}) Zhang-Li torque \cite{ZhangLi2004}, and \emph{(ii)} spin torque from the spin dependent Seebeck effect \cite{slachter2010thermally}. However, neither of these torques follows the experimentally observed $\mathbf{H} \cdot \mathbf{I} > 0$ symmetry, as their sign would not depend on the direction of the applied current. 

%
The observed symmetry instead strongly suggests that spin-orbit torque (SOT) is the main source of anti-damping. 
Recent results obtained in ferromagnetic/oxide heterostructures suggest that 
SOT can result from non-equilibrium spin accumulation due to inversion symmetry breaking and enhanced spin-orbit coupling at the interfaces\cite{Hals2010epl, kurebayashi2014antidamping,Kim2017prb}. 
In principle, intrinsic anomalous \cite{Das2017prb,Bose2018prappl} and spin Hall effects in Py \cite{Azevedo2015prb,Tsukahara2014prb} and interfacial Rashba effect \cite{Miron2011} might all contribute to the SOT in our system. 
In fact, our micromagnetic simulations, assuming a spin Hall/Rashba-like SOT, require a spin Hall angle of about 0.13 to reproduce the auto-oscillation threshold currents observed in the experiments, which is in the range of values reported for Py/oxide stacks\cite{manchon2018current}.

Besides the fundamental interest of generating, controlling, and optimizing SOT in single ferromagnetic layers, \emph{e.g.}~using ultra-low damping metals and different oxide interfaces, our results will have a direct impact on a wide range of 
applications. A magnetic tunnel junction (MTJ) in the nano-constriction region only requires an additional ferromagnetic layer, which will allow for separate optimization of the SOT drive and the MTJ based high-power read-out. Arrays of SOT driven nano-constrictions can also be used for oscillator based neuromorphic computing \cite{Torrejon2017nature}. In the sub-threshold regime, the intrinsic damping in magnonic crystals can be greatly reduced, solving their issues with high transmission losses \cite{KruglyakJPD2010}. 

\section*{Acknowledgments:} This work was supported by the European Research Council (ERC) under the European Community’s Seventh Framework Programme (FP/2007-2013)/ERC Grant 307144 “MUSTANG.” This work was also supported by the Swedish Research Council (VR), the Swedish Foundation for Strategic Research (SSF), and the Knut and Alice Wallenberg Foundation, and the Wenner-Gren Foundation.
\newline

\clearpage

\clearpage

\end{document}